\documentclass[11pt]{article}
\usepackage{amsmath,amssymb}
\begin{document}
\title{{\bf The Uniqueness Problem of Sequence Product on Operator Effect Algebra ${\cal E}(H)$}\thanks{E-mail: wjd@zju.edu.cn}}
\author {{Liu Weihua,\,\,  Wu Junde}\\
{\small\it Department of Mathematics, Zhejiang University, Hangzhou
310027, P. R. China}}

\date{}
\maketitle

{\small\it \noindent Abstract. A quantum effect is an operator on a
complex Hilbert space $H$ that satisfies $0\leq A\leq I$. We denote
the set of all quantum effects by ${\cal E}(H)$. In this paper we
prove, Theorem 4.3, on the theory of sequential product on ${\cal
E}(H)$ which shows, in fact, that there are sequential products on
${\cal E}(H)$ which are not of the generalized L\"{u}ders form. This
result answers a Gudder's open problem negatively.}


\vskip 0.1 in

\noindent {\bf 1. Introduction}

\vskip 0.1 in

\noindent If a quantum-mechanical system $\cal S$ is represented in
the usual way by a complex Hilbert space $H$, then a self-adjoint
operator $A$ on $H$ such that $0\leq A\leq I$ is called the {\it
quantum effect} on $H$ ([1, 2]). Quantum effects represent yes-no
measurements that may be unsharp. The set of quantum effects on $H$
is denoted by ${\cal E}(H)$. The subset ${\cal P}(H)$ of ${\cal
E}(H)$ consisting of orthogonal projections represents sharp yes-no
measurements. Let ${\cal T}(H)$ be the set of trace class operators
on $H$ and ${\cal S}(H)$ the set of density operators, i.e., the
trace class positive operators on $H$ of unit trace, which represent
the states of quantum system. An {\it operation} is a positive
linear mapping $\Phi: {\cal T}(H)\rightarrow {\cal T}(H)$ such that
for each $T\in {\cal S}(H)$, $0\leq tr[\Phi (T)]\leq 1$ ([3-5]).
Each operation $\Phi$ can define a unique quantum effect $B$ such
that for each  $T\in {\cal T}(H)$, $tr[\Phi(T)]=tr[TB]$.

\noindent Let ${\cal B}(H)$ be the set of bounded linear operators
on $H$, the dual mapping $\Phi^*: {\cal B}(H)\rightarrow {\cal
B}(H)$ of an operation $\Phi$ is defined by the relation
$tr[T\Phi^*(A) ]=tr[\Phi(T)A ]$, $A\in{\cal B}(H), T\in {\cal T}(H)$
([4]). The effect $B$ defined by an operation $\Phi$ satisfies that
$B=\Phi^*(I)$ ([5]).

\noindent For each $P\in {\cal P}(H)$ is associated a so-called
L\"{u}ders operation $\Phi_L^P: T\rightarrow PTP$, its dual is
$(\Phi_L^P)^*(A)=PAP$ and the corresponding quantum effect is
$(\Phi_L^P)^*(I)=P$. These operations arise in the context of ideal
measurements. Moreover, each quantum effect $B\in {\cal E}(H)$ gives
 to a general L\"{u}ders operation $\Phi_L^B: T\rightarrow
B^{\frac{1}{2}}TB^{\frac{1}{2}}$ and $B$ is recovered as
$(\Phi_L^B)^*(I)=B$ as well.

\noindent Let $\Phi_1, \Phi_2$ be two operations. The composition
$\Phi_2\circ \Phi_1$ is a new operation, called a sequential
operation as it is obtained by performing first $\Phi_1$ and then
$\Phi_2$. In general, $\Phi_2\circ \Phi_1\neq \Phi_1\circ \Phi_2$.
Note that for any two quantum effects $B, C\in {\cal E}(H)$ we have
$(\Phi_L^C\circ \Phi_L^B)^*(I)=B^{\frac{1}{2}}CB^{\frac{1}{2}}$
$([5, P_{26-27}])$. It shows that the new quantum effect
$B^{\frac{1}{2}}CB^{\frac{1}{2}}$ yielded by $B$ and $C$ has
important physics meaning. Professor Gudder called it the sequential
product of $B$ and $C$, and denoted it by $B\circ C$. It represents
the quantum effect produced by fist measuring $A$ then measuring $B$
([6-8]). This sequential product has also been generalized to an
algebraic structure called a {\it sequential effect algebra} ([7]).

\vskip 0.1 in

\noindent Now, we introduce the abstract sequential product on
${\cal E}(H)$ as following:

\noindent Let $\circ$ be a binary operation on ${\cal E}(H)$, i.e.,
$\circ$: ${\cal E}(H)\times {\cal E}(H)\rightarrow {\cal E}(H)$, if
it satisfies:

\noindent (S1). The map $B\rightarrow A\circ B$ is additive for each
$A\in {\cal E}(H)$, that is, if $B+C\leq I$, then

$(A\circ B)+(A\circ C)\leq I$ and $(A\circ B)+(A\circ C)=A\circ
(B+C)$.

\noindent (S2). $I\circ A=A$ for all $A\in {\cal E}(H)$.

\noindent (S3). If $A\circ B=0$, then $A\circ B=B\circ A$.

\noindent (S4). If $A\circ B=B\circ A$, then $A\circ
(I-B)=(I-B)\circ A$ and $A\circ (B\circ C)=(A\circ B)\circ C$ for

all $C\in {\cal E}(H)$.

\noindent (S5). If $C\circ A=A\circ C$, $C\circ B=B\circ C$, then
$C\circ (A\circ B)=(A\circ B)\circ C$ and

$C\circ (A+B)=(A+B)\circ C$ whenever $A+B\leq I$.

\noindent If ${\cal E}(H)$ has a binary operation $\circ$ satisfying
conditions (S1)-(S5), then $({\cal E}(H), 0, I, \circ)$ is called a
sequential operator effect algebra. Professor Gudder showed that for
any two quantum effects $B$ and $C$, the operation $\circ$ defined
by $B\circ C=B^{\frac{1}{2}}CB^{\frac{1}{2}}$ satisfies conditions
(S1)-(S5), and so is a sequential product of ${\cal E}(H)$, which we
call the generalized L\"{u}ders form. In 2005, Professor Gudder
presented 25 open problems about the general sequential effect
algebras. The second problem is:

\noindent {\bf Problem 1.1 ([9])}. Is $B\circ
C=B^{\frac{1}{2}}CB^{\frac{1}{2}}$ the only sequential product on
${\cal E}(H)$?

\noindent As we see the five properties are base on the measurement
logics and the the uniqueness property has been asked many times in
Gudder's paper. In this paper, we construct a new sequential product
on ${\cal E}(H)$ which differs from the generalized L\"{u}ders form,
thus, we answer the open problem negatively.

\vskip 0.1 in

\noindent {\bf 2. Sequential Product on ${\cal E}(H)$}

\vskip 0.1 in

\noindent In this section, we study some abstract properties of
sequential product $\circ $ on ${\cal E}(H)$. For convenience, we
introduce the following notations: If $A, B\in{\cal E}(H)$, we say
that $A\oplus B$ is defined if and only if $A+B\leq I$ and define
$A\oplus B=A+B$; if $A\circ B=B\circ A$, we denote $A|B$ .

\vskip 0.1 in

\noindent {\bf Lemma 2.1.} If $A, B \in {\cal E}(H), a\in [0,1]$,
then
$$A\circ (aB)=a(A\circ B).$$

\noindent Proof. It is clear that for $a=1$, the conclusion is true.
If $a>0$ is a rational number, i.e., $a=\frac{m}{n}$, where $n, m$
are positive integer, it follows from $\bigoplus\limits_{i=1}^{n}
(A\circ {\frac{1}{n}B})=A\circ B$ that $A\circ
(\frac{1}{n}B)=\frac{1}{n}(A\circ B)$,  thus, $A\circ
(\frac{m}{n}B)=\bigoplus\limits_{i=1}^{m}
A\circ(\frac{1}{n}B)=\frac{m}{n}(A\circ B)$. If $a\in [0,1]$ is not
a rational number, then for each $q=\frac{m}{n}>a$ we have $q(A\circ
B)=A\circ (qB)=A\circ[(q-a)B]+A\circ(aB)\geq A\circ(aB)$, so $
q(A\circ B)\geq A\circ(aB)$. Let $q\rightarrow a$ we have $a(A\circ
B)\geq A\circ(aB)$. Similarly, we can get that $A\circ(aB)\geq
a(A\circ B)$ by taking $q=\frac{m}{n}<a$. So $A\circ(aB)=a(A\circ
B)$. Moreover, it follows from the proof process that for $a=0$ the
conclusion is also true.

\vskip 0.1 in

\noindent {\bf Lemma 2.2} ([9], Theorem 3.4 (i)). Let $A\in {\cal
E}(H)$ and $E\in {\cal P}(H)$. If $A\leq E$, then $A|E$ and $E\circ
A=A$.

\vskip 0.1 in

\noindent {\bf Lemma 2.3.} If $a\in [0,1]$, $E\in {\cal P}(H)$, then
$aI|E$ and $(aI)\circ E=E\circ(aI)=aE$.

\noindent Proof. Since $aE\leq E$, so $aE|E$ and $E\circ E=E$ by
Lemma 2.2, it follows from $E=E\circ I=(E\circ E) \oplus
(E\circ(I-E))=E\oplus (E\circ (I-E))$ that $ E\circ (I-E)=0$, note
that $E\circ (a(I-E))\leq E\circ (I-E)=0$, so $E\circ (a(I-E))=0$,
thus, it follows from (S3) that $E|a(I-E)$, moreover, by (S5) we
have $E|a(I-E)\oplus aE=aI$, so, it follows from Lemma 2.1 and Lemma
2.2 that $(aI)\circ E=E\circ(aI)=a(E\circ I)=aE$.

\vskip 0.1 in

\noindent {\bf Lemma 2.4.} If $E, F\in {\cal P}(H), E\leq F$ and
$0\leq a\leq 1$, then $E|aF$ and $E\circ(aF)=aE$.

\noindent Proof. It follows from $E\leq F$ that $I-E\geq I-F\geq
a(I-F)$, by Lemma 2.2 and Lemma 2.3, we have $I-E|a(I-F)$ and
$I-E|(1-a)I$, thus, $I-E|a(I-F)\oplus(1-a)I=I-aF$, it follows from
(S4) that $E|I-aF$ and so by (S4) again that $E|aF$, moreover, by
Lemma 2.1 and Lemma 2.2, we have $(aF)\circ E=E\circ(aF)=a(E\circ
F)=aE$.

\vskip 0.1 in

\noindent {\bf Lemma 2.5.} If $E\in {\cal P}(H), A\in {\cal E}(H),
0\leq a \leq 1 $ and $A\leq E$, then $aE|A$, and $(aE)\circ
A=A\circ(aE)=aA$.

\noindent Proof. It follows from Lemma 2.2 that $A|E$, so by (S4) we
have $A|I-E$. Since $A\circ E=A=A\circ I=A\circ E \oplus A\circ
(I-E)$, so $A\circ(I-E)=0$. Note that $A\circ (a(I-E))\leq
A\circ(I-E)$, we have $A\circ (a(I-E))=0$, so $A|a(I-E)$.

\vskip 0.1 in

\noindent Let $\{E_{\lambda}\}$ be the identity resolution of $A$
and denote
$$A_n=\sum\limits_{i=0}^{2^n-1}\frac{i}{2^n}(E_{\frac{i+1}{2^n}}-E_{\frac{i}{2^n}}),$$
$$B_n=\sum\limits_{i=1}^{2^n}\frac{i}{2^n}(E_{\frac{i}{2^n}}-E_{\frac{i-1}{2^n}}).$$
Note that $A\in \varepsilon(H)$, so $E_{\lambda}=0$ when $\lambda <
0$ and $E_{\lambda}=I$ when $1\leq \lambda$. Moreover, for each
$n\in \mathbb{N}$, $A_n\leq A_{n+1}$, $B_{n+1}\leq B_{n}$, and when
$n\rightarrow \infty$, $\|A_n-A\|\rightarrow 0, \|B_n-A\|\rightarrow
0$ ([10]).

Let $0\leq b \leq 1$. Then it follows from Lemma 2.1 and Lemma 2.3
that
$$(bI)\circ A_n=\sum
\limits_{i=1}^{2^n-1}(bI)\circ(\frac{i}{2^n})(E_{\frac{i+1}{2^n}}-E_{\frac{i}{2^n}})$$$$=\sum
\limits_{i=1}^{2^n-1}(\frac{ib}{2^n})(E_{\frac{i+1}{2^n}}-E_{\frac{i}{2^n}})=bA_n$$
and
$$(bI)\circ B_n=bB_n.$$ Note that $ A\geq A_n$, so $(bI)\circ
A\geq (bI)\circ A_n=bA_n$. Let $n\rightarrow \infty$. Then
$(bI)\circ A\geq bA$, do the same with$\{B_n\}$, we get $(bI)\circ
A\leq bA$, so $(bI)\circ A= bA=A\circ(bI)$. That is $A|bI$ for each
$0\leq b\leq 1$, in particular, $A|(1-a)I$. Thus, it follows from
$A|(1-a)I+a(I-E)$ that $A|I-aE$, by (S4) we have $A|aE$, Hence,
$(aE)\circ A=A\circ(aE)=a(A\circ E)=aA$.

\vskip 0.1 in

\noindent {\bf  Lemma 2.6.} Let $0\leq a\leq 1$ and $A, B\in {\cal
E}(H)$. Then $$(aA)\circ B=A\circ (aB)=a(A\circ B).$$

\noindent Proof. It follows from Lemma 2.5 that $(aA)\circ B=(A\circ
(aI))\circ B=A\circ((aI)\circ B)=A\circ (aB)=a(A\circ B)$.

\vskip 0.1 in

\noindent Lemma 2.6 showed that we can write $a(A\circ B)$ for
$(aA)\circ B$ and $A\circ(aB)$.

\vskip 0.1 in

\noindent In order to obtain our main result in this section, we
need to extent $\circ :{\cal E}(H)\times{\cal E}(H)\rightarrow {\cal
E}(H)$ to ${\cal E}(H)\times {\cal S}(H)\rightarrow {\cal S}(H)$,
where ${\cal S}(H)$ is the set of bounded linear self-adjoint
operators on $H$.

\vskip 0.1 in

\noindent Let $B\in {\cal E}(H)$, $A\in {\cal S}^+(H)$. Then there
exists a number $M>0$ such that $\frac{A}{M} \in {\cal E}(H).$ Now
we define
$$B\circ A=M(B\circ \frac{A}{M}).$$ If there is another positive number
$M'$ such that $\frac{A}{M'}\in {\cal E}(H)$, without losing
generality, we assume that $M\leq M'$, then $M'(B\circ
\frac{A}{M'})=M'(B\circ(\frac{M}{M'}\frac{A}{M}))
=M'(\frac{M}{M'}(B\circ\frac{A}{M}))=M(B\circ \frac{A}{M})$, this
showed that $B\circ A$ is well defined for each bounded linear
positive operator $A$ on $H$.

\noindent In general, if $A\in {\cal S}(H)$, we can express $A$ as
$A_1-A_2$, where $A_1,A_2 $ are two bounded linear positive
operators on $H$  ([10]). Now we define
$$B\circ A=B\circ A_1-B\circ A_2.$$ If $A_1'-A_2'$ is another
expression of $A$ with the above properties, then
$A_1+A_2'=A_1'+A_2=K$ is a bounded linear positive operator on $H$.
If take positive real number $M$ such that $\frac{K}{M}\in{\cal
E}(H)$, then $B\circ( A_1+A_2')=M(
B\circ(\frac{A_1}{M}+\frac{A_2'}{M}))=M(
B\circ\frac{A_1}{M})+M(B\circ\frac{A_2'}{M})=B\circ A_1+B\circ
A_2'$. Similarly, $B\circ(A_1'+A_2)=B\circ A_1'+B\circ A_2$. Thus,
it follows from $B\circ A_1'+B\circ A_2=B\circ A_1+B\circ A_2'$,
$B\circ A_1-B\circ A_2=B\circ A_1'-B\circ A_2'$. This showed that
$\circ $ is well defined on ${\cal E}(H)\times S(H)$.

\vskip 0.1 in

\noindent From the above discussion we can easily prove the
following important result:

\vskip 0.1 in

\noindent {\bf Theorem 2.7.} If $B\in{\cal E}(H)$, $A_1, A_2\in
S(H)$ and $a\in \mathbb{R}$, then we have $$B\circ(A_1+A_2)=B\circ
A_1+B\circ A_2,\,\, B\circ(aA_1)=a(B\circ A_1).$$

\vskip 0.2 in

\noindent {\bf 3. Sequential Product on ${\cal E}(H)$ with
dim$(H)=2$}

\vskip 0.2 in

\noindent In this section, we suppose that dim$(H)=2$. Now, we
explore the key idea of constructing our sequential product.

\vskip 0.1 in

\noindent {\bf  Lemma 3.1.} If $E\in {\cal P}(H), B\in {\cal E}(H)$,
then $E\circ B=EBE.$

\noindent Proof. Since $E$ is a orthogonal projection on ${\cal
E}(H)$ with dim$(H)=2$, so there exists a normal basis $\{e_1,
e_2\}$ of $H$ such that $E(e_i)=\lambda_i e_i$, where $\lambda_i\in
\{0, 1\}$, $i=1, 2$. If $\lambda_i=0, i=1, 2$, then $E=0$, if
$\lambda_i=1, i=1, 2$, then $E=I$. It is clear that for $E=0$ or
$E=I$, the conclusion is true. Without losing generality, we now
suppose that $\lambda_1=1$ and $\lambda_2=0$, i.e., $(E(e_1),
E(e_2))=(e_1, e_2)\left(
\begin{array}{cc}
1&0\\
0&0\\
\end{array}
\right).$ Let $B\in S(H)$. Then we have $\begin{array}{cc} (B(e_1),
B(e_2))=(e_1, e_2)\left(
\begin{array}{cc}
x&y\\
\bar{y}&z\\
\end{array}
\right)\end{array}$, where $x, z\in \mathbb{R}$ ([10]). Now we
define two linear operators $X$ and $Z$ on $H$ satisfy that
$$(X(e_1), X(e_2))=(e_1, e_2)\left(
\begin{array}{cc}
x&0\\
0&0\\
\end{array}
\right)$$ and $$(Z(e_1), Z(e_2))=(e_1, e_2)\left(
\begin{array}{cc}
0&0\\
0&z\\
\end{array}
\right).$$ Then $X=xE, Z=z(I-E)\in {\cal E}(H)$ and it follows from
(S1) and Lemma 2.2 that $E\circ X=X$ and $E\circ Z=0$. Denote
$$(E\circ B(e_1), E\circ B(e_2))=(e_1, e_2)\left(
\begin{array}{cc}
f(x,y,z)&g(x.y.z)\\
\overline{g(x.y.z)}&h(x,y,z)\\
\end{array}
\right).$$ Since $S(H)$ is a real linear space and by Theorem 2.7
that $B\rightarrow E\circ B$ is a real linear map of
$S(H)\rightarrow S(H)$, so $f,g$ and $h$ are real linear maps of
vector $(x, y, z)$ and $f$ and $g$ are real-valued functions of $(x,
y, z)$, thus, function $f(x,y,z)$ must have the form ([10]):
$f(x,y,z)=kx+lz+n(y+\bar{y})+im(y-\bar{y}),$ where $k,l,m,n\in R$.
Let $B=X$ and $B=Z$, respectively, it follows from $E\circ X=X$ and
$E\circ Z=0$ that $l=0, k=1$, so
$f(x,y,z)=x+n(y+\bar{y})+mi(y-\bar{y})$. Note that when $B\in {\cal
S}^+(H)$, $E\circ B$ should be a positive operator, so when
$x,z\geq0$ and $xz-|y|^2\geq0$, we have $f(x,y,z)\geq0$. Take $y\in
R$, then $f(x,y,z)=x+2ny$. Thus, when $x,z\geq 0$, $y\in R$ and
$xz-y^2\geq 0$, $f(x,y,z)=x+2ny\geq 0$. If $n\neq 0$, take
$y=-\frac{1}{n}$, $x=1$, $z=\frac{1}{n^2}$, then we have $f<0$, this
is a contradiction and so $n=0$. Similarly, if $m\neq 0$, take
$y=-\frac{i}{m}$, $x=1$, $z=\frac{1}{m^2}$, we will get $f<0$, this
is also a contradiction and so $m=0$. Thus, we have $f(x,y,z)=x$.

\noindent Moreover, note that $E\circ ((I-E)\circ B)=(E\circ
(I-E))\circ B=0\circ B=0=((I-E)\circ E))\circ B=(I-E)\circ(E\circ
B)$, as above, we may prove that $((I-E)\circ(E\circ B)(e_1),
(I-E)\circ(E\circ B)(e_2))=(e_1, e_2)\left(
\begin{array}{cc}
0&0\\
0&h(x,y,z)\\
\end{array}
\right)=(e_1, e_2)\left(
\begin{array}{cc}
0&0\\
0&0\\
\end{array}
\right)$, thus $h(x,y,z)=0$. For each $y\in \mathbb{C}$, take $x=1$,
$z=|y|^2$, then $B$ is a positive operator, so $E\circ B$ is also a
positive operator, thus we have $fh-|g|^2\geq 0$. It follows from
$h=0$ that $g=0$, so $E\circ B=X=EBE$.

\vskip 0.1 in

\noindent {\bf Corollary 3.2.} Let $E\in {\cal P}(H), a\in [0,1]$
and $A=aE$. Then for each $B\in {\cal E}(H)$, $$A\circ B=(aE)\circ
B=a(E\circ
B)=a(EBE)=a^{\frac{1}{2}}EBa^{\frac{1}{2}}E=A^{\frac{1}{2}}BA^{\frac{1}{2}}.$$

\vskip 0.1 in

\noindent Now, we prove the following important result:

\vskip 0.1 in

\noindent {\bf Theorem 3.2.} Let $H$ be a complex Hilbert space with
dim$(H)=2$, $A, B\in {\cal E}(H)$. If $\{e_1, e_2\}$ is a normal
basis of $H$ such that $(A(e_1), A(e_2))=(e_1, e_2)\left(
\begin{array}{cc}
a^2&0\\
0&b^2\\
\end{array}
\right)$ and $(B(e_1), B(e_2))=(e_1, e_2)\left(
\begin{array}{cc}
x&y\\
\bar{y}&z\\
\end{array}
\right)$, then there exists a $\theta\in \mathbb{R}$ such that
$$(A\circ B(e_1), A\circ B(e_2))=(e_1, e_2)\left(
\begin{array}{cc}
a^2x&abe^{i\theta}y\\
abe^{-i\theta}\bar{y}&b^2z\\
\end{array}
\right).$$

\noindent Proof. Let $\{e_1, e_2\}$ be a normal basis of $H$ such
that $(A(e_1), A(e_2))=(e_1, e_2)\left(
\begin{array}{cc}
a^2&0\\
0&b^2\\
\end{array}
\right)$ and $(B(e_1), B(e_2))=(e_1, e_2)\left(
\begin{array}{cc}
x&y\\
\bar{y}&z\\
\end{array}
\right)$, where $0\leq a, b\leq 1$, $0\leq x, 0\leq z, 0\leq
xz-|y|^2$. Now we define a linear operator $E$ on $H$ such that
$(E(e_1), E(e_2))=(e_1, e_2)\left(
\begin{array}{cc}
1&0\\
0&0\\
\end{array}
\right)$, then $E\in{\cal P}(H)$. By Corollary 3.2, we can suppose
$a, b\in (0, 1]$ and $a\neq b$. Thus, $A=a^2E+b^2(I-E)$. Denote
$(A\circ B(e_1), A\circ B(e_2))=(e_1, e_2)\left(
\begin{array}{cc}
f(x,y,z)&g(x,y,z)\\
\overline{g(x,y,z)}&h(x,y,z)\\
\end{array}
\right)$, where $f, g, h$ are real linear functions with respect to
$(x, y, z)\in \mathbb{R}\times \mathbb{C}\times \mathbb{R}$ and $f,
h$ take values in $\mathbb{R}$. Since $E\circ(A\circ B)=(E\circ
A)\circ B)=(E\circ (a^2E+b^2(I-E)))\circ B=a^2(E\circ B)$, we have
$f(x,y,z)=a^2x$. Similarly, we have also $h(x, y, z)=b^2z$.
Moreover, since $E|E, E|(I-E)$, by (S5), we have $E|A$, so by (S4),
we have $(I-E)|A$, thus, $A\circ(xE)=xa^2E$, $A\circ
z(I-E)=zb^2(I-E)$, this showed that $g$ is independent of $x$ and
$z$, so $g(x,y,z)=\alpha y$, where $\alpha\in C$. On the other hand,
if $B\in {\cal S}(H)$ is a positive operator, then $A\circ B$ is
also a positive operator, so for each positive number $x$ and $z$,
and each complex number $y$, when $xz-|y|^2\geq 0$, we have
$a^2b^2xz-|\alpha y|^2\geq 0$. Let $x=1$, $z=|y|^2$. Then we get
that
$$a^2b^2-|\alpha|^2\geq 0. \eqno (1).$$

\vskip 0.1 in

\noindent Let $B, C$ be two positive operators. We show that if both
$B\leq C$ and $C\leq B$ are not true, then both $A\circ B\leq A\circ
C$ and $A\circ C\leq A\circ B$ are also not true. In fact, let
$D=b^2E+a^2(I-E)$. Then $A|b^2E+a^2(I-E)=D$ and $A\circ
D=A\circ(b^2E+a^2(I-E))=a^2b^2I$. So if $A\circ B\leq A\circ C$,
then $D\circ(A\circ B )\leq D\circ(A\circ C)$. But $D\circ(A\circ
B)=(D\circ A)\circ B=a^2b^2I\circ B=a^2b^2B\leq D\circ(A\circ
C)=a^2b^2C$, thus we will have $B\leq C$, this is a contradiction.
So $A\circ B\leq A\circ C$ is not true. Similarly, we have $A\circ
C\leq A\circ B$ is also not true.

\noindent Let $y\in\mathbb{C}$, $y\neq 0$, $\epsilon$ be a positive
number satisfy that $a^2|y|-\epsilon>0$. If we define $(B(e_1),
B(e_2))=(e_1, e_2)\left(
\begin{array}{cc}
|y|&y\\
\bar{y}&|y|\\
\end{array}
\right)$ and $(C(e_1, C(e_2))=(e_1, e_2)\left(
\begin{array}{cc}
\epsilon &0\\
0&0\\
\end{array}
\right)$, then $B, C\in{\cal E}(H)$, $B\leq C$ and $C\leq B$ are
both not true. Thus we have both $A\circ B\leq A\circ C$ and $A\circ
B\leq A\circ C$ are also not true, i.e., the self-adjoint operator
$A\circ B-A\circ C$ is not positive operator. Note that $((A\circ
B-A\circ C)(e_1), (A\circ B-A\circ C)(e_2))=(e_1, e_2)\left(
\begin{array}{cc}
a^2|y|-\epsilon&\alpha y\\
\overline{\alpha y}&b^2|y|\\
\end{array}
\right)$, and $a^2|y|-\epsilon>0$, $b^2|y|>0$, so we have
$b^2(a^2|y|-\epsilon)|y|-|\alpha y|^2<0$. Let $\epsilon\rightarrow
0$, we get that $|\alpha y|^2\geq b^2a^2|y|^2$. Thus, we have
$$|\alpha|^2\geq b^2a^2. \eqno (2)$$ It follows from (1)
and (2) that $|\alpha|^2=a^2b^2.$ So $|\alpha|=ab$ and
$\alpha=abe^{i\theta}$.

\vskip 0.2 in

\noindent {\bf 4. A New Sequential Product on ${\cal E}(H)$}

\vskip 0.2 in

\noindent Theorem 3.2 motivated us to construct the new sequential
product on ${\cal E}(H)$. First, we need the following:

\noindent For each $A\in {\cal E}(H)$, denote $R(A)=\{Ax, x\in H\}$,
$N(A)=\{x, x\in H, Ax=0\}$, $P_0$ and $P_1$ be the orthogonal
projections on $\overline {R(A)}$ and $N(A)$, respectively. It
follows from $A\in {\cal E}(H)$ that $N(A)=N(A^{1/2})$, so
$R(A)=R(A^{1/2})$. Moreover, $P_0(H)\bot P_1(H)$ and $H=P_0(H)\oplus
P_1(H)$ ([10]).

\noindent Denote $f_z(u)$ be the complex-valued Borel function
defined on $[0, 1]$, where $f_z(u)=\exp z(\ln u)$ if $u\in (0, 1]$
and $f_z(0)=0$. Now, we define
$$A^i=f_i(A),\,\, A^{-i}=f_{-i}(A).$$ It is easily to show that $||A^i||\leq 1$, $||A^{-i}||\leq 1$ and $$(A^i)^*=A^{-i}, A^iA^{-i}=
A^{-i}A^{i}=P_0.$$

\vskip 0.1 in

\noindent {\bf Theorem 4.1.} Let $H$ be a complex Hilbert space and
$A, B\in {\cal E}(H)$. If we define $A\circ
B=A^{1/2}A^{i}BA^{-i}A^{1/2}$, then $\circ$ satisfies the conditions
(S1)-(S3).

\noindent Proof. If $A, B\in {\cal E}(H)$, note that $||A^i||\leq 1$
and $||A^{-i}||\leq 1$, we have
$$\| A\circ B\|=\|A^{1/2}A^{i}BA^{-i}A^{1/2}\|\leq
\|A^{1/2}\|\|A^{i}\|\|B\|\|A^{-i}\|\|A^{1/2}\|\leq 1$$ and
$$<A^{1/2}A^{i}BA^{-i}A^{1/2}x,x>=\|B^{1/2}A^{-i}A^{1/2}x\|\geq 0$$
for all $x\in H$, so $A\circ B=A^{1/2}A^{i}BA^{-i}A^{1/2}$ is a
binary operation on ${\cal E}(H)$. Moreover, it is clear that the
map $B\rightarrow A\circ B$ is additive for each $A\in {\cal E}(H)$,
so the operation $\circ$ satisfies (S1).

\noindent It follows from $I\circ A=I^{1/2}I^{i}AI^{-i}I^{1/2}=A$
that $\circ$ satisfies (S2).

\noindent If $A\circ B=A^{1/2}A^{i}BA^{-i}A^{1/2}=0$, now, we
represent $A$ and $B$ on $H=P_0(H)\oplus P_1(H)$ by $\left(
\begin{array}{cc}
A_1 &0\\
0 &0\\
\end{array}
\right)$  and $\left(
\begin{array}{cc}
B_1 &B_2\\
B_3 &B_4\\
\end{array}
\right)$, then $$A\circ B=\left(
\begin{array}{cc}
A_1^{1/2}A_1^{i}B_1A_1^{-i}A_1^{1/2} &0\\
0 &0\\
\end{array}
\right)=0,$$ so we have $A_1^{1/2}A_1^{i}B_1A_1^{-i}A_1^{1/2}=0$ on
$P_0(H)$, i.e., $(A_1^{1/2}A_1^{i}B_1A_1^{-i}A_1^{1/2}x,x)=0$ for
each $x\in P_0(H)$. Note that $R(A)=R(A^{1/2})$ and  $A^i$ is a
unitary operator on $P_0(H)$, so $R(A^{1/2})$ is dense in $P_0(H)$,
thus for each $y\in P_0(H)$, there is a sequence $\{z_n\}\subseteq
R(A^{1/2})$ such that $z_n\rightarrow A^iy$, so there is a sequence
$\{x_n\}\subseteq H$ such that $A^{1/2}x_n=z_n\rightarrow A^iy$. Let
$x_n=y_n+u_n$, where $y_n\in P_0(H), u_n\in P_1(H)$. Then
$A^{1/2}x_n=A^{1/2}y_n$. Thus, there is a sequence $\{y_n\}$ in
$P_0(H)$ such that $A^{1/2}y_n=z_n\rightarrow A^iy$. Note that $A^i$
is a unitary operator on $P_0(H)$, so we have
$A^{-i}A^{1/2}y_n\rightarrow y$. But,
$$\|B_1^{1/2}A_1^{-i}A_1^{1/2}y_n\|=(A_1^{1/2}A_1^{i}B_1A_1^{-i}A_1^{1/2}y_n, y_n)=0,$$
so $B_1^{1/2}y=0$ for each $y\in P_0(H)$, that is, $B_1^{1/2}=0$.
Since $B\in {\cal E}(H)$, so $B_2=0, B_3=0$, thus we have $B=\left(
\begin{array}{cc}
0 &0\\
0 &B_4\\
\end{array}
\right)$, so $B\circ A=B^{1/2}B^{i}AB^{-i}B^{1/2}=0=A\circ B$. This
showed that $\circ$ satisfies (S3).

\vskip 0.1 in

\noindent {\bf Theorem 4.2.} Let $H$ be a complex Hilbert space with
dim$(H)<\infty$, $A, B\in {\cal E}(H)$. If we define $A\circ
B=A^{1/2}A^{i}BA^{-i}A^{1/2}$, then $A\circ
B=A^{1/2}A^{i}BA^{-i}A^{1/2}=B\circ A=B^{1/2}B^{i}AB^{-i}B^{1/2}$ if
and only if $AB=BA$.

\noindent {\bf Proof.} Firstly, it is obvious that if $AB=BA$, then
$A\circ B=A^{1/2}A^{i}BA^{-i}A^{1/2}=B\circ
A=B^{1/2}B^{i}AB^{-i}B^{1/2}$. Now, if $A\circ
B=A^{1/2}A^{i}BA^{-i}A^{1/2}=B\circ A=B^{1/2}B^{i}AB^{-i}B^{1/2}$,
we show that $AB=BA$. Note that $A\in {\cal E}(H)$ and
dim$(H)<\infty$, so $A$ has the form $\sum\limits_{i=1}^{n} a_iE_i$,
where $\sum\limits_{k=1}^{n} E_k=I$, $a_k\geq 0$, $E_k\in {\cal
P}(H)$, $a_k\neq a_l$, $E_kE_l=0$ for all $k, l=1, 2, \cdots, n,
k\neq l$. Without losing generality, we suppose that $0\leq
a_1<\cdots<a_n$, then $0\leq|a_1^{1/2}f_i(a_1)|<\cdots
<|a_n^{1/2}f_i(a_n)|$ since $a_k^{1/2}=|a_k^{1/2}f_i(a_k)|$. It
follows from the operator theory that $A^{1/2}=\sum\limits_{k=1}^{n}
a_k^{1/2}E_k$ and $f_i(A)=A^i=\sum\limits_{k=1}^{n} f_i(a_k)E_k$,
$f_{-i}(A)=A^{-i}=\sum\limits_{k=1}^{n} f_{-i}(a_k)E_k$ ([10]). Note
that $A^{1/2}A^{i}BA^{-i}A^{1/2}=B^{1/2}B^{i}AB^{-i}B^{1/2}$, so for
each $x\in H$, $(A^{1/2}A^{i}BA^{-i}A^{1/2}x,
x)=(B^{1/2}B^{i}AB^{-i}B^{1/2}x, x)$, thus we have
$$\|B^{1/2}A^{-i}A^{1/2}x\|=\|A^{1/2}B^{-i}B^{1/2}x\|.\eqno (3)$$ Take $x\in
E_n(H)$, then $A^{1/2}A^{-i}x=A^{-i}A^{1/2}x=a_n^{1/2}f_{-i}(a_n)x$,
note that $|a_nf_{-i}(a_n)|=|a_nf_{i}(a_n)|=|a_n|$, $\overline
{R(B)}=\overline {R(B^{1/2})}$ and $B^{-i}$ is a unitary operator on
$\overline {R(B)}$ and $B^{-i}B^{1/2}=B^{1/2}B^{-i}$, we have
$$\|A^{1/2}B^{1/2}B^{-i}x\|^2=\|\sum\limits_{k=1}^{n}
a_k^{1/2}E_kB^{1/2}B^{-i}x\|^2=$$$$\sum\limits_{k=1}^{n}
a_k\|E_kB^{1/2}B^{-i}x\|^2\leq \sum\limits_{k=1}^{n}
a_n\|E_kB^{1/2}B^{-i}x\|^2=$$$$a_n||B^{1/2}B^{-i}x\|^2=||a_n^{1/2}B^{-i}B^{1/2}x\|^2=$$$$||a_n^{1/2}B^{1/2}x\|^2=||B^{1/2}A^{1/2}A^{-i}x\|^2.$$
Thus, it follows from equation (3), $B^{-i}B^{1/2}=B^{1/2}B^{-i}$,
$A^{-i}A^{1/2}=A^{1/2}A^{-i}$ and $0\leq a_1<\cdots<a_n$ that for
each $k<n$, we have $E_kB^{1/2}B^{-i}x=0$, so $B^{1/2}B^{-i}x\in
E_n(H)$. Thus we have $E_nB^{1/2}B^{-i}E_n=B^{1/2-i}E_n$. This
showed that $B^{1/2}B^{-i}$ has the matrix form $\left(
\begin{array}{cc}
C &D\\
0 &K\\
\end{array}
\right)$ on $H=E_n(H)\oplus(I-E_n)(H)$, where $C\in {\cal B}(E_n(H),
E_n(H))$,
    $D\in {\cal B}((I-E_n)(H), E_n(H)), K\in {\cal B}((I-E_n)(H), (I-E_n)(H))$. Note that $B\in
{\cal E}(H)$, $B$ has the form $\sum\limits_{k=1}^{m} b_kF_k$, and
$B^{1/2}B^{-i}=\sum\limits_{k=1}^{m}b^{1/2}f_{-i}(b_k)F_k$, where
$\sum\limits_{k=1}^{m} F_k=I$, $b_k\geq 0$, $F_k\in{\cal P}(H)$,
$b_k\neq b_l$, $F_kF_l=0$ for all $k, l=1, 2,\cdots, m, k\neq l$.
Now we define a polynomial $$G_k(z)=\prod\limits_{j\neq
k}(z-b_j^{1/2}f_{-i}(b_j))/\prod\limits_{j\neq
    k}(b_k^{1/2}f_{-i}(b_j)-b_j^{1/2}f_{-i}(b_j))$$ on $\mathbb{C}$.
It is easily to show that for each $1\leq k\leq m$,
$G_k(B^{1/2}B^{-i})=F_k$. Note that $B^{1/2}B^{-i}$ has the
up-triangulate form, so $G_k(B^{1/2}B^{-i})$ has also the
up-triangulate form. But $F_k$ is a self-adjoint operator, so $F_k$
has the diagonal matrix form on $E_n(H)\oplus(I-E_n)(H)$. This
implies that $F_k$ commutes with $E_n$ for each $k$, so $B$ commutes
with $E_n$. Denote $A_0=A-a_nE_n$, then we still have $A_0\circ
B=B\circ A_0$ as discussed before, thus we get that $B$ commutes
with $E_{n-1}$. Continuously, we will have that $B$ commutes with
all $E_k$ and so with A. In this case we have $A\circ B=AB$.

\vskip 0.1 in

\noindent Our main result is:

\vskip 0.1 in

\noindent {\bf Theorem 4.3.} Let $H$ be a complex Hilbert space with
$dim(H)< \infty $ and $A, B\in {\cal E}(H)$. If we define $A\circ
B=A^{1/2}A^{i}BA^{-i}A^{1/2}$, then $\circ$ is a sequential product
on ${\cal E}(H)$.

\noindent {\bf Proof.} By Theorem 4.1, we only need to prove that
$\circ$ satisfies (S4) and (S5). In fact, if $A|B$, i.e., $A\circ
B=A^{1/2}A^{i}BA^{-i}A^{1/2}=B\circ A=B^{1/2}B^{i}AB^{-i}B^{1/2}$,
then it follows from Theorem 4.2 that $A$ commutes with $B$ and of
course $I-B$, so $A|I-B$. If $C\in{\cal E}(H)$, we have
$$A\circ (B\circ
C)=A^{\frac{1}{2}}A^iB^{\frac{1}{2}}B^iCB^{-i}B^{\frac{1}{2}}A^{-i}A^{\frac{1}{2}}$$$$=A^{\frac{1}{2}}B^{\frac{1}{2}}A^iB^iCA^{-i}B^{-i}A^{\frac{1}{2}}B^{\frac{1}{2}}$$$$=(AB)^{\frac{1}{2}}(AB)^iC(AB)^{-i}(AB)^{\frac{1}{2}}
$$$$=(AB)\circ C=(A\circ B)\circ C.$$ So (S4) is satisfied.

\noindent Moreover, if $C|B$ and $C|A$, then $C(AB)=ACB=(AB)C$,
$C(A\oplus B)=(B+A)C$, so it is easily to prove that $C(A\circ
B)=(A\circ B)C$, thus, by Theorem 4.2, we have $C|A\circ B$ and
$C|(A\oplus B)$ whenever $A\oplus B$ is defined, this showed that
(S5) is hold.

\vskip 0.1 in

\noindent By using Theorem 4.3 we can prove the following corollary:

\vskip 0.1 in

\noindent {\bf Corollary 4.4.} Let $H$ be a complex Hilbert space
with dim$(H)=2$, $A, B\in{\cal E}(H)$. Take a normal basis $\{e_1,
e_2\}$ of $H$ such that $(A(e_1), A(e_2))=(e_1, e_2)\left(
\begin{array}{cc}
a^2&0\\
0&b^2\\
\end{array}
\right)$ and $(B(e_1), B(e_2))=(e_1, e_2)\left(
\begin{array}{cc}
x&y\\
\bar{y}&z\\
\end{array}
\right)$. If when $a, b>0$, define $$((A\circ B)(e_1), (A\circ
B)(e_2))=(e_1, e_2)\left(
\begin{array}{cc}
a^2x&abe^{i\theta}y\\
abe^{-i\theta}\bar{y}&b^2z\\
\end{array}
\right),$$ where $\theta =\ln a^2-\ln b^2$; when $a>0, b=0$, define
$$((A\circ B)(e_1), (A\circ B)(e_2))=(e_1, e_2)\left(
\begin{array}{cc}
a^2x&0\\
0&0\\
\end{array}
\right),$$ when $a=0, b>0$, define
$$((A\circ B)(e_1), (A\circ B)(e_2))=(e_1, e_2)\left(
\begin{array}{cc}
0&0\\
0&b^2z\\
\end{array}
\right),$$ then $\circ$ is a sequential product of ${\cal E}(H)$.

\vskip 0.1 in

\noindent {\bf Remark 1.} In conclusion, we construct a new
sequential product $A\circ
B=A^{\frac{1}{2}}A^iBA^{-i}A^{\frac{1}{2}}$ on $\varepsilon(H)$ with
dim$(H)<\infty$, which is different from the generalized L\"{u}ders
form $A^{\frac{1}{2}}BA^{\frac{1}{2}}$. In this proof we can also
get a more general one $A\circ
B=A^{\frac{1}{2}}A^{ti}BA^{-ti}A^{\frac{1}{2}}$ for $t\in R$. It
indicates that with the measurement rule (S1)-(S5), there can be a
time parameter $t$ to describe the phase change. In particular, if
dim$(H)=2$, $A\in{\cal E}(H)$ and $\{e_1, e_2\}$ is a normal basis
of $H$ such that $(A(e_1), A(e_2))=(e_1, e_2)\left(
\begin{array}{cc}
a^2&0\\
0&b^2\\
\end{array}
\right)$, then when $a>0, b>0$ and $a\neq b$, Corollary 4.4 showed
that $\theta =(\ln a^2-\ln b^2)t$ can be used to describe the
phase-changed phenomena of quantum effect $A\circ B$. As the proof
showed, it is the only form that the sequential product can be. This
is much more important in physics. \noindent

\vskip 0.1 in

\noindent {\bf Remark 2.} As we knew, in the quantum computation and
quantum information theory, if $(A_i)^n_{i=1}\subseteq {\cal B}(H)$
satisfying $\sum\limits_{i=1}^{n}A_iA_i^*=I$, then the operators
$(A_i)^n_{i=1}$ are called the operational elements of the quantum
operation
 $U:{\cal T}(H)\rightarrow {\cal T}(H)$ defined by $$U(\rho)=\sum\limits_{i=1}^{n}A_i\rho A_i^*
 ,$$ where ${\cal T}(H)$ is the set of trace class operators. Any trace preserving, normal, completely positive map has the above
 form. This is very important in  describing dynamics, measurements, quantum channels, quantum interactions, and quantum
 error, correcting codes, etc. If $(A_i)^n_{i=1}$ is a set of quantum effects
with $\sum\limits_{i=1}^{n}A_i=I$,
 then the transformation $U'(\rho)=\sum\limits_{j=1}^{n}A_j^{\frac{1}{2}}A_j^{ti}\rho A_j^{-ti}A_j^{\frac{1}{2}}
 $ is a well defined quantum operation since $\sum\limits_{j=1}^{n}A_j^{\frac{1}{2}}A_j^{ti} A_j^{-ti}A_j^{\frac{1}{2}}
 =\sum\limits_{i=1}^{n}A_i=I$. So this new sequential product
 yields a natural and interesting quantum operation.

\vskip 0.1 in

\noindent {\bf Remark 3.} Theorem 4.3 indicates that the conditions
(S1)-(S5) of sequential product of ${\cal E}(H)$ are not sufficient
to characterize the generalized L\"{u}ders form
$A^{\frac{1}{2}}BA^{\frac{1}{2}}$ of $A$ and $B$. Recently,
Professor Gudder presented a characterization of the sequential
product of ${\cal E}(H)$ is the generalized L\"{u}ders form ([11]).

\vskip 0.2 in

\noindent {\bf Acknowledgement.} The authors wish to express their
thanks to the referees for their valuable comments and suggestions.
In particular, their comments motivated the authors to prove Theorem
4.3 for any finite dimensional Hilbert spaces. This project is
supported by Natural Science Foundations of China (10771191 and
10471124).

\vskip 0.2 in

\noindent {\bf References }

\vskip 0.2 in

\noindent [1] Ludwig G 1983 {\it Foundations of Quantum Mechanics
(I-II)} (Springer, New York)

\noindent [2] Ludwig G 1986 {\it An Axiomatic Basis for Quantum
Mechanics (II)} (Springer, New York)

\noindent [3] Kraus K 1983 {\it Effects and Operations}
(Springer-Verlag, Beilin)

\noindent [4] Davies E B 1976 {\it Quantum Theory of Open Systems}
(Academic Press, London)

\noindent [5] Busch P, Grabowski M and Lahti P J 1999 {\it
Operational Quantum Physics} (Springer-Verlag, Beijing Word
Publishing Corporation)

\noindent [6] Gudder S,  Nagy G 2001 J. Math. Phys. 42 5212

\noindent [7] Gudder S, Greechie R 2002 Rep Math. Phys. 49 87

\noindent [8] Gheondea A, Gudder S 2004 Proc. Am. Math. Soc. 132 503

\noindent [9] Gudder S 2005 Inter. J. Theory. Physi. 44 2199

\noindent [10] Kadison R V, Ringrose J R 1983 {\it Fundamentals of
the Theory of Operator algebra} (Springer, New York)

\noindent [11] Gudder S,  Latremoliere F 2008 J. Math. Phys. 49
052106

\end{document}